\documentclass[twocolumn,showpacs,amssymb,prb]{revtex4}
\usepackage{graphicx,amstext}

\begin{document}

\title{Origin of Colossal Dielectric Response of Pr$_{0.6}$Ca$_{0.4}$MnO$_{3}$}

\author{N. Bi$\check{s}$kup}
 \email{biskup@icmm.csic.es}
\author{A. de Andr$\acute{e}$s}
\author{J.L. Martinez}
\affiliation{Instituto de Ciencia de Materiales, CSIC, Cantoblanco
28049 Madrid, Spain}
\author{C. Perca}
 \affiliation {Laboratoire de Physico-Chimie de l'Etat Solide, Universit$\acute{e}$ Paris XI, 91405 Orsay Cedex - France}

\date{\today}

\begin{abstract}

We report the detailed study of dielectric response of
Pr$_{0.6}$Ca$_{0.4}$MnO$_{3}$ (PCMO), member of manganite family
showing colossal magnetoresistance. Measurements have been performed
on four polycrystalline samples and four single crystals, allowing
us to compare and extract the essence of dielectric response in the
material. High frequency dielectric function is found to be
$\varepsilon$$_{0}$=30, as expected for the perovskite material.
Dielectric relaxation is found in frequency window of 20Hz-1MHz at
temperatures of 50-200K that yields to colossal low-frequency
dielectric function, i.e. static dielectric constant. Static
dielectric constant is always colossal, but varies considerably in
different samples from $\varepsilon$$_{0}$=10$^{3}$ until 10$^{5}$.
The measured data can be simulated very well by blocking (surface
barrier) capacitance in series with sample resistance. This
indicates that the large dielectric constant in PCMO arises from the
Schottky barriers at electrical contacts. Measurements in magnetic
field and with d.c. bias support this interpretation. Colossal
magnetocapacitance observed in the title compound is thus attributed to extrinsic effects. Weak anomaly
at the charge ordering temperature can also be attributed to
interplay of sample and contact resistance. We comment our results
in the framework of related studies by other groups.

\end{abstract}

\maketitle 

\textbf{Introduction}

The typical value of dielectric constant $\varepsilon$(0) in solids
is of the order 1-10. Exceptions are ferroelectrics with
$\varepsilon$(0)$\approx$10$^{4}$ and (charge/spin) density waves
materials (CDW/SDW) with $\varepsilon$(0)$\approx$10$^{7}$-10$^{9}$.
In the former the large dielectric response is a consequence of
charge polarization due to ferroelectric displacement of central ion
in the unit cell. In the latter, large polarization is achieved by
local displacement of electron condensate in density wave. But both
of them are unsuitable for applications: ferroelectrics due to
limited temperature and frequency range around ferroelectric
transition and CDW/SDW materials due to inapplicably low
temperatures where density waves occur.

It is therefore understandable that discovery of room temperature
frequency independent "colossal" dielectric constant of complex
perovskite compound CaCu$_{3}$Ti$_{4}$O$_{12}$ (CCTO)
\cite{Subramanian} sparked the interest in new materials that might
not be limited by frequency and temperature. At room temperature
CCTO has high dielectric constant
($\varepsilon$(0)$\approx$10$^{4}$-10$^{5}$) that was confirmed in
ceramic samples, single crystals and thin films. Theoretical
modeling has excluded the possibility of intrinsic origin of high
$\varepsilon$(0) \cite{Cohen}. These studies conclude that the
internal inhomogeneities are in the origin of the effect. It is
suspected that those inhomogeneities arise from crystal twinning or
some internal domain boundaries. On the other hand, some authors
interpret high dielectric response as an artifact coming from
Schottky effect at the electrode contacts \cite{LunkenheimerCCTO}.
In order to solve this dispute, it is useful to study dielectric
response in other materials that are known to be inhomogeneous.
Manganites are excellent candidates for this purpose.

The family of manganites has attracted a widespread attention of
scientific community in the last 15 years due to their "colossal"
magnetoresistance \cite{Dagottorev}. The magnetoresistive effects in
some compositions reach the factor of 10$^{6}$, which essentially
means magnetic field induced insulator-metal transitions. In order
to understand such colossal effects, the concept of phase separation
has emerged. The ground state of certain manganite compounds
\cite{Uehara, Fath} is proved to be inhomogeneous, consisting
typically of metallic clusters in an insulating/semiconducting
matrix. Such a separation of phases (metallic and insulating) is
believed to be the cause of many unusual phenomena, including
colossal magnetoresistance. The purpose of this work is to
investigate if the phase separating boundaries can contribute to the
dielectric response.

The general formula for manganites is R$_{1-x}$A$_{x}$MnO$_{3}$
where R stands for rear earth (La, Pr) and A for any divalent atom
(Ca, Sr). Among various manganites, Pr$_{1-x}$Ca$_{x}$MnO$_{3}$ is
unique, showing insulating behaviour over the whole composition (x)
range due to its narrow bandwidth of 3d conducting e$_{g}$ electrons
\cite{Tomioka}. The title compound Pr$_{0.6}$Ca$_{0.4}$MnO$_{3}$
falls in the range 0.3$<$x$<$0.75 where the ground state is charge
ordered antiferromagnetic insulator \cite{Jirak, Yoshizawa}. Charge
ordering refers to ordering of manganese ions that can be in
Mn$^{3+}$ or Mn$^{4+}$ valence: at low temperatures these ions order
into a superstructure forming stripes of Mn$^{3+}$ and Mn$^{4+}$
ions. Magnetization measurements in the title compound show two
transitions at T=240 and T=180K that are believed to be charge
(T$_{CO}$) and antiferromagnetic ordering (T$_{AF}$), respectively.
Insulator-metal transition in title compound can be induced by
magnetic field \cite{Tomioka}, electric current \cite{current
switch}, pressure \cite{pressure} or X-rays \cite{X-ray switch}.
These and other studies \cite{optical,thermal,neutron} converged
around the idea of phase separation involving ferromagnetic metallic
droplets coalescing and enabling the current percolation through the
insulating matrix. The origin of phase separation lies in existence
of structural inhomogeneities (clusters) associated with charge
ordering \cite{Radaelli,Kajimoto}. Recent report of nanoscale
competition in charge ordered LCMO \cite{clusters} concludes that
even the same phases (charge ordered insulator) form clusters with
different orientation of CO stripes. The formation of clusters
generally precedes structural or magnetic transitions \cite{Dagotto}
arising at temperature T$^{*}$$>$T$_{C}$ (in our case
T$_{C}$=T$_{CO}$). PCMO system is thus ideal to study dielectric
response of the inhomogeneous (clustered) system. Initial reports of
giant dielectric response in Pr$_{0.67}$Ca$_{0.33}$MnO$_{3}$ (x=1/3)
appeared in year 1999 \cite{Rivas} and the most recent one in
2004\cite{Mercone}. In the latter case it is suggested that it
arises from CDW orderings or phase separation inhomogeneities. We
have performed detailed study on similar system (PCMO with x=3/8) in
order to resolve the origin of apparent colossal dielectric response
in PCMO.

After this introduction, we give an overview of experimental
methods. In results we report on four topics: resistive
characterization, dielectric response in temperature, influence of
magnetic field and influence of d.c. bias. Discussion is following
each of these measurement reports. In Summary we bring the
conclusions and comment our findings in the light of related
reports.\vspace{0.2cm}

\textbf{Experimental}

We have measured 4 polycrystalline (PC) samples (P1-P4) and 4 single
crystals (SC, S1-S4) which enables us to extract the data inherent
to the material itself and test and compare the findings. Single
crystal samples are plates that are cut from the single rod. A Laue
pattern taken on the growth direction of the single crystal
indicated the coincidence with the [001] direction, inside 15
degree, or even less. Single crystal measurements for samples S1-S3
are with current contacts in this [001] direction while contacts for
S4 are in perpendicular direction. The nominal composition of
polycrystals (Pr$_{0.6}$Ca$_{0.4}$MnO$_{3}$) differs slightly from
the single crystals (Pr$_{0.625}$Ca$_{0.375}$MnO$_{3}$) but their
characterization shows almost identical behaviour (Figure 1).
Samples are first characterized magnetically (SQUID), later
electrically (four-contact configuration) and then prepared for the
capacitance measurements (two-contact configuration). Care was taken
to reduce the parasite capacitances of measurement system below 3pF
in order to have measurable signal even in the high frequency limit
($\varepsilon$$_{HF}$). Dielectric measurements are done with
Quadtech LCR-meter, model 1920. The voltage applied was always at
low limit of 20mV in order to minimize the effects of
voltage-current nonlinearity. The frequency range covered by this
instrument lies between 20Hz and 1MHz. At low temperatures the
relaxation times for colossal dielectric response drop below our
frequency window. Here we measured dielectric constant/capacitance
through the time dependent charging effect by sourcemeter Keithley
2410.\vspace{0.2cm}

\textbf{Results and discussion }\vspace{0.2cm}

\emph{Resistive characterization}\vspace{0.1cm}

 Figure 1 shows the
resistive characterization of representative polycrystalline (P1)
and single crystal (S1) samples.
\begin{figure}
\centering\includegraphics[clip,scale=0.43]{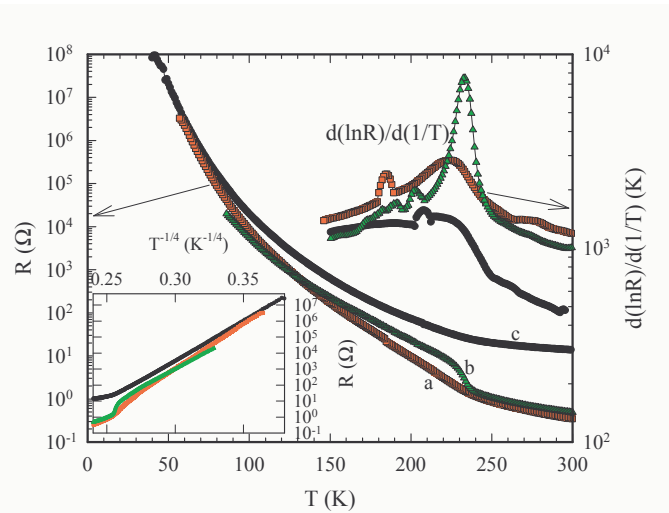}
\caption{ (Color online) Temperature dependent resistance of
polycrystal P1 Pr$_{0.6}$Ca$_{0.4}$MnO$_3$ (a, squares) and single
crystal S1 Pr$_{5/8}$Ca$_{3/8}$MnO$_3$ (b, triangles). Circles (c)
stand for two contact (capacitance) configuration of P1 sample. In
the right inset is d(lnR)/d(1/T) indicating T$_{CO}$. In the left
inset is resistance over T$^{-1/4}$ indicating three dimensional
variable range hopping.} \label{Fig1}
\end{figure}
The charge ordering transition, interpreted as the peak in the
derivative d(lnR)/d(1/T), appears at T$_{CO}$=225 and 235K,
respectively. These values are very close, especially in the light
of "broad" peak in polycrystalline sample. Magnetization data give
identical value for both cases of T$_{CO}$=235K. The
antiferromagnetic ordering at 175 K is detectable only in the
magnetization measurements. Small peaks in derivative curves at
T$\leq$200K are consequence of single-point cracks in resistivity
measurements and therefore not related to antiferromagnetic
ordering.  The insulating behavior below T$_{CO}$ is governed by
three-dimensional (d=3) variable range hopping \cite{Mott} where
electrical conductivity follows

$$\sigma_{DC}=\sigma_{0}e^{-(\frac{T_{0}}{T})^{\frac{1}{1+d}}}\eqno{(1)}$$

Further, all data from resistive
measurements (resistivity in polycrystal and two perpendicular
directions in single crystals, as well as measurements in magnetic
field) show no sign of anisotropy. This enables us to treat
identically the dielectric response in polycrystalline and single
crystal samples. From Figure 1 we can also estimate the influence of
contact resistances in two-probe measurements, which, as expected,
becomes negligible at low temperatures.\vspace{0.2cm}

\emph{Dielectric response and its temperature
dependence}\vspace{0.1cm}

The most instructive presentation of dielectric data is given in the
terms of frequency dependent dielectric permittivity/capacitance
\cite{Jonscher} - the method that is followed here. Dielectric data
are collected measuring real (G) and imaginary (B) part of
electrical admittance. Capacitance is directly related to dielectric
constant as C=$\varepsilon$$_{0}$$\varepsilon$(s/l) and in this
report we present data in this form. Figures 2 shows the typical
frequency response of PCMO system: Fig 2a shows real and Fig 2b imaginary part of
complex capacitance C*=C'+iC".
\begin{figure}
\centering\includegraphics[clip,scale=0.43]{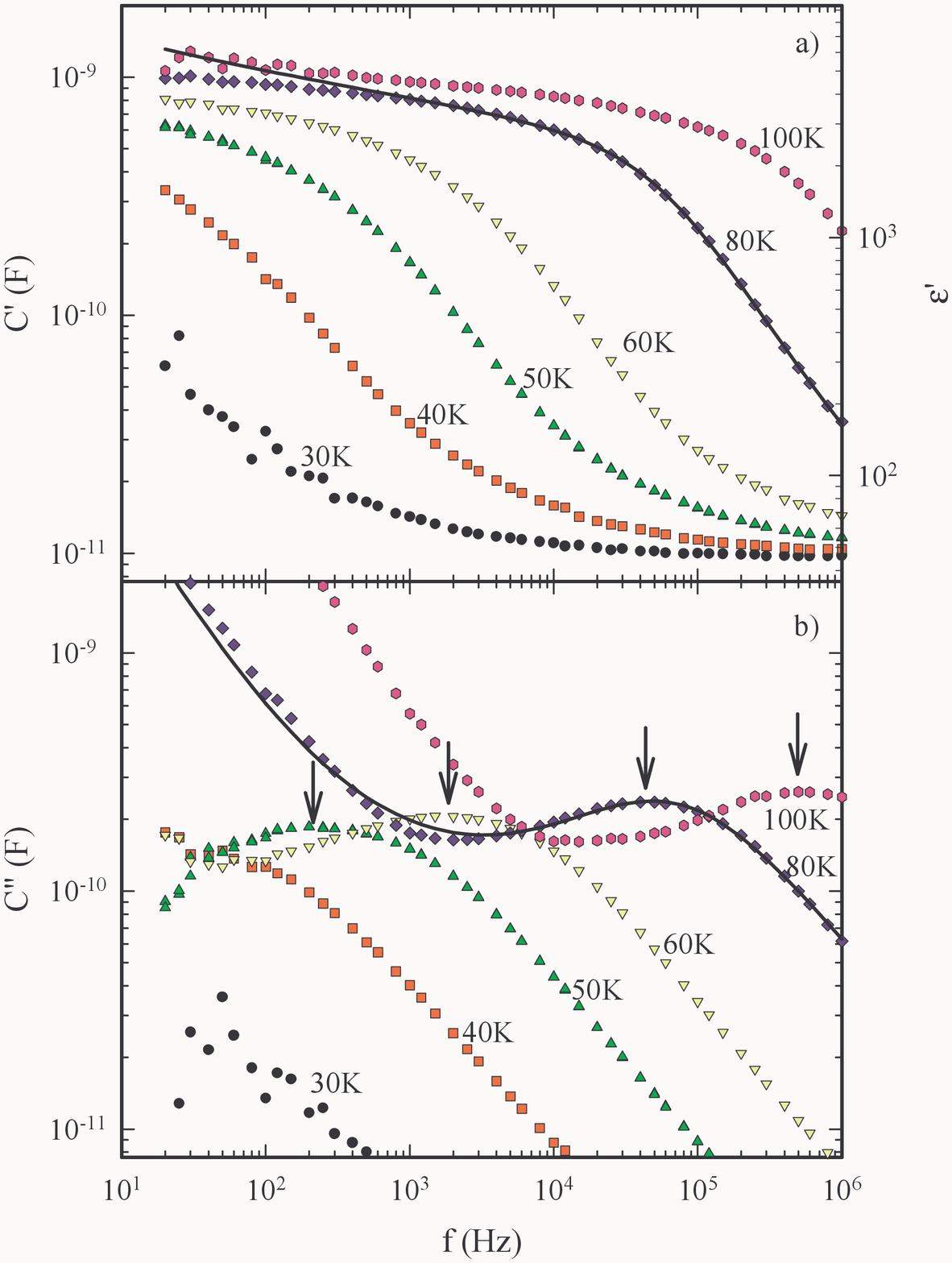}
\caption{ (Color online) Dielectric relaxation for polycrystal P1 at
temperatures given in the figures. a) C'($\omega$). On the right scale is value of
dielectric function. b) C"($\omega$). Arrows denote loss peaks. Line is the fit to 80K data as defined in text.}
\label{Fig2}
\end{figure}
The data in figures 2 correspond to P1 polycrystal but represent
very well the response of all eight samples. The low-frequency
relaxation shown in Fig. 2 is
very reminiscent to Debye relaxation, having well defined 'loss
peak' in $\varepsilon$". Phenomenologically, such relaxation is
given by:

$$\varepsilon(\omega)=\varepsilon_{HF}+[\varepsilon(0)-\varepsilon_{HF}]\frac{1}{1-i\omega\tau_{0}}\eqno{(2)}$$

where $\tau$$_{0}$ denotes characteristic low-frequency relaxation
time $\tau$$_{0}$=1/$\omega_{0}$. $\varepsilon$(0) is dielectric
constant and $\varepsilon_{HF}$ stands for high frequency dielectric
function, i.e. at $\omega\gg\omega$$_0$=1/$\tau$$_0$. In the limit
$\varepsilon$$_{HF}\rightarrow$0 this expression corresponds to
'Debye' equivalent circuit consisting of resistance R coupled
serially to capacitance C. Relaxation time
$\tau$$_{0}$ in such equivalent circuit is given by $\tau$$_{0}$=RC
and at frequencies above relaxation
($\omega$$>$$\omega$$_0$=1/$\tau$$_{0}$, real and imaginary part of
complex 'capacitance' follow $\omega$$^{-2}$ and $\omega$$^{-1}$
dependence, respectively. Note however that corresponding exponents
in Fig. 2 are lower than in this ideal (Debye) case. This is usually
interpreted in the terms of distribution of relaxation times.

\begin{figure}
\centering\includegraphics[clip,scale=0.43]{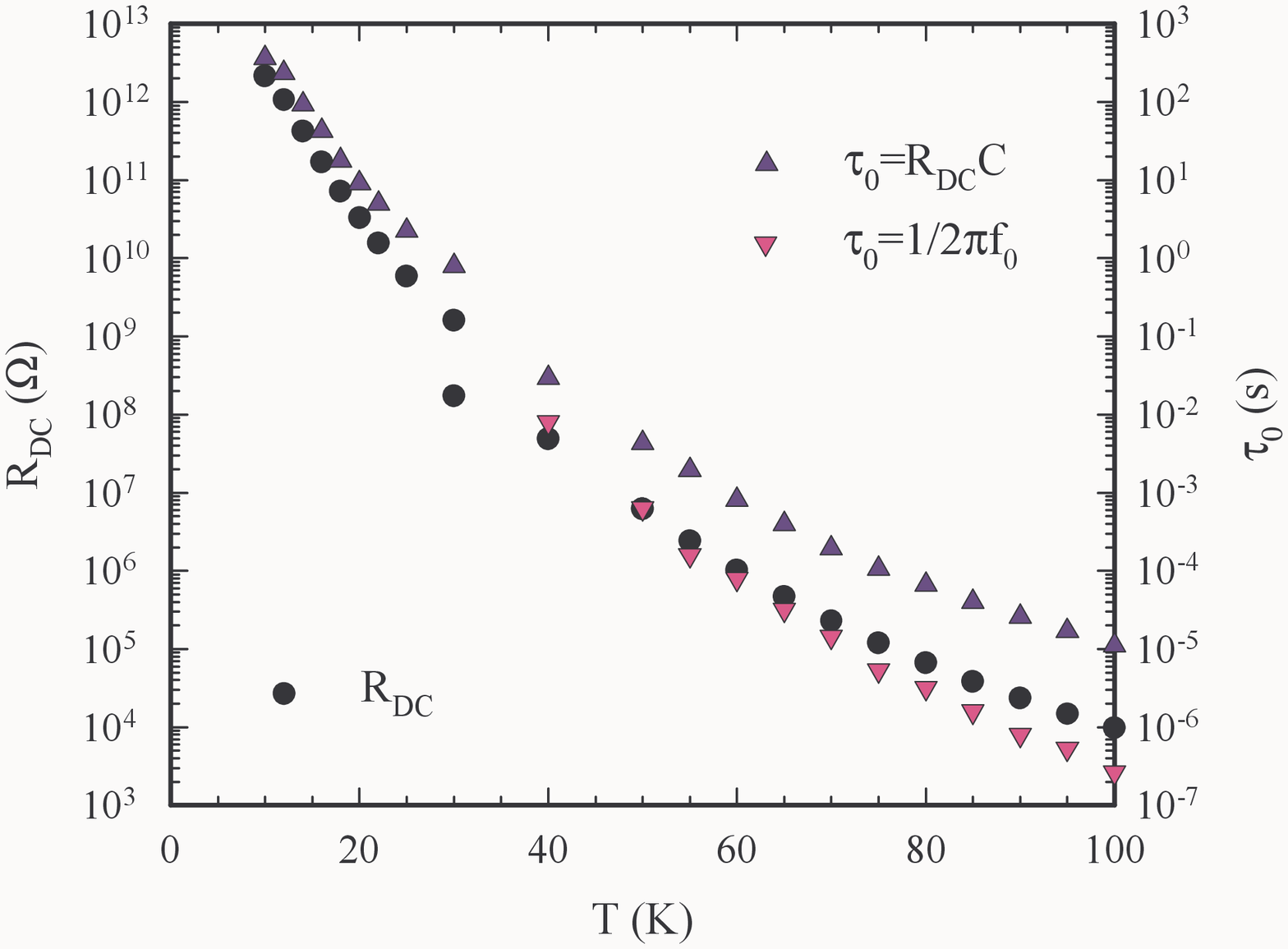} \caption{
(Color online) Fig. 3 Relaxation time $\tau_0$ as a function of
temperature (triangles - right axis). See text for point assignment. On the left
axis is samples resistance (solid circles).} \label{Fig3}
\end{figure}
Figure 3 shows the resistance R of sample P1 (left axis) plotted
together with 'Debye' relaxation time $\tau$$_0$=1/$\omega_0$ (right
axis) for temperatures below 100K. All data are from two contact
capacitance measurements. Solid circles denote R$_{DC}$ data taken
by LCR-meter. At T$\leq$30K relaxation falls below our frequency
window. At these temperatures capacitance C and resistance R are
measured through time dependence of charging process. Voltage at the
electrodes is built up according to
V(t)=V$_{0}$(1-e$^{-\frac{t}{RC}}$), which extends our temperature
range for capacitance measurements down to T=10K. Triangles pointing
up are simply product of resistance and capacitance $\tau_0$=R$_{DC}$C,
while those pointing down are taken from the loss peak frequency
$\tau_0$=1/2$\pi$f$_{0}$, as indicated in figure 2b. One can see
that $\tau_0$ follows fairly well the temperature behavior of
resistance. This shouldn't be surprising if dielectric screening
arises from the same carriers that contribute to electric
conductivity. The discrepancy of two "definitions" of $\tau_0$
decreases with decreasing temperature and tends to diminish at very
low temperatures where contact resistance becomes negligible
comparing to intrinsic sample resistance entering into $\tau_0$=R$_{DC}$C.
This will be an important argument in considerations below.

As we can see from figures 2 and 3, our dielectric data can be
represented quite well by Debye equivalent circuit consisting of
capacitance in series with resistance \cite{Jonscher}.
\begin{figure}
\centering\includegraphics[clip,scale=0.45]{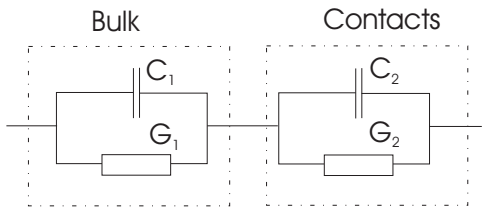}
\caption{ Equivalent circuit for combination of bulk and surface
components.} \label{Fig4}
\end{figure}
High frequency capacitance, when parasite capacitance is
substracted, gives $\varepsilon_{HF}$$\approx$30, as expected for
perovskites. Low frequency capacitance of sample P1 gives colossal
values of $\varepsilon$(0)$\simeq$5000. In order to verify these
data, we have measured four polycrystalline samples and four single
crystals and, although varying considerably, dielectric constant
always shows "colossal" values above 10$^{3}$. Dielectric constant
and relevant data for all 8 samples are presented in Table 1.
\begin{table}[!h]
\begin{tabular}{|c|c|c|c|c|c|c|c|}
  \hline
  $sample$ & $area$  & $thickness$  & $contact$ & $R_{RT}$ & $C(0)$ & $\varepsilon(0)$ \\
  &$(mm^{2}$)&$(mm)$&$type$&($\Omega)$&$(nF)$&\\
  \hline
  $P1$ & $12$ & $0.5$ & $Ag paint$ & $10$ & $1$ & $5\cdot10^{3}$ \\
  $P2$ & $13.2$ & $1.2$ & $film Au$ & $9$ & $2$ & $2\cdot10^{4}$ \\
  $P3$ & $7$ & $1.2$ & $film Au$ & $24$ & $9$ & $1.7\cdot10^{5}$ \\
  $P4$ & $4.1$ & $0.63$ & $Ag paint$ & $20$ & $0.12$ & $3\cdot10^{3}$ \\
  $S1$ & $9$ & $0.25$ & $Ag paint$ & $2$ & $0.4$ & $1.3\cdot10^{3}$ \\
  $S2$ & $9$ & $0.65$ & $Ag paint$ & $3$ & $0.25$ & $2\cdot10^{3}$ \\
  $S3$ & $7.2$ & $0.58$ & $Ag paint$ & $54$ & $0.14$ & $1.3\cdot10^{3}$ \\
     &     &      & $film Au$ & $35$ & $5$ & $4.5\cdot10^{4}$ \\
  $S4$ & $2.9$ & $1.4$ &  $film Au$  & $98$ & $6.5$ &$3.6\cdot10^{5}$\\
  \hline
\end{tabular}
  \caption{ Some relevant parameters of eight samples in this study. All resistances are from two-contact measurements} \label{TablART}
\end{table}
Value of dielectric constant is deduced by extrapolation of flat
part of $\varepsilon$' (i.e. at $\omega$$\ll$$\omega_0$) toward zero
frequency. One can see that lowest dielectric constant (and
capacitance) have samples with contacts made directly with silver
paint. Samples with preevaporated gold contacts show much higher
dielectric constant: $\varepsilon$(0)$\geq$10$^{4}$ despite higher
(or just because of it!) contact resistances. This finding is even
emphasized by large difference of C(0) on the same sample (S3) for
two different types of contacts. It can be also seen that
capacitance doesn't depend (at least significantly) on geometrical
factors. All of the above suggest that dielectric response in PCMO
is governed by contacts.

   In Figure 4 we present a typical equivalent circuit that represents both bulk and
surface (blocking) capacitances. Both of them are parallely
accompanied by their corresponding resistances or, as noted in the
Figure 4, by conductances G$_{i}$=1/R$_{i}$. Zero frequency capacitance C(0) for such a circuit is given by:

$$C(0)=\frac{G_{1}^{2}C_{2}+G_{2}^{2}C_{1}}{(G_{1}+G_{2})^{2}}\eqno{(3)}$$

In the high temperature limit we assume G$_{2}$$\ll$G$_{1}$ that
yields to C(0)=C$_{2}$ and in low temperature limit
G$_{2}$$\gg$G$_{1}$ yielding C(0)=C$_{1}$. Between these two limits
we have complex interplay of conductances G$_1$ and G$_2$ that
determinates capacitance C(0) and relaxation time $\tau$$_0$. Let's
now assume that bulk capacitance C$_1$ equals to high frequency
limit of capacitance in Fig. 2, i.e. C$_1$=10pF. Surface or contact
capacitance is then the one of low frequency limit, i.e. C$_2$=1nF.
As we saw from Fig.1, contact resistance at low temperatures
diminishes comparing to bulk resistance. Capacitance is in this, low
temperature limit given by
C=(G$_1$$^2$/G$_2$$^2$)C$_2$=(R$_2$$^2$/R$_1$$^2$)C$_2$. In this way
the diminishing contribution of contact resistance can explain the
decrease of capacitance at low temperatures (as is evident from Fig
2). Relaxation time $\tau_0$ in low temperature limit
(G$_2$$>$G$_1$) is given predominantly by G$_1$, like in the typical
Debye case (R$_1$=1/G$_1$ in series with C$_2$). This gives the same
temperature dependence of $\tau$$_0$ as shown in Fig.3. Note further
that two definitions of $\tau$$_0$ plotted in Fig.3 differ more at
high temperatures and converge toward low temperatures. This
convergence illustrates diminishing contribution of contact
resistance R$_2$ in R=R$_1$+R$_2$ since real relaxation time
$\tau_0$ (measured by dielectric relaxation) is defined by bulk
resistance $\tau_0$=R$_1$C$_2$ and not by overall resistance
R=R$_1$+R$_2$. Our low-frequency relaxation thus seems to come from
contact capacitances. Finally, to successfully fit our data to the
combination of bulk and surface
 elements like in Fig.4, we assumed that element C$_2$ is not ideal but the
universal capacitance. This means that C$_2$ is frequency dependent
(C$_2$=B(i$\omega$)$^{n-1}$) which actually simulates the distribution
of different contact capacitances coming from irregularities at the
contact interfaces. Such an assumption is necessary to successfully
fit broadened relaxation (n$<$1) from Fig. 2. Fit for T=80K is shown
as a solid line in Fig.2.

In Table 1 we have listed values of capacitances C(0) for our eight
samples. These are the values at room temperature (RT) that are
either measured directly or deduced from the low temperature
measurements. Namely, due to small resistance of PCMO samples at
room temperature, dielectric measurements are impeded by both
inductances and sensitivity of measuring instrument, especially at
low frequencies. Thus, we were able to record RT capacitances
directly only in samples with high C(0). Figure 5 presents C(0)
values for four of our samples up to (and above) room temperature.
One can see once again that the capacitance increases with
temperature and becomes nearly temperature independent
\begin{figure}
\centering\includegraphics[clip,scale=0.43]{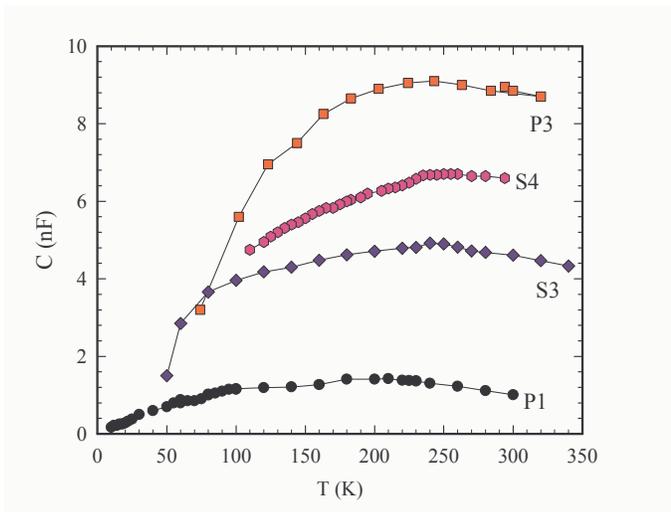}
\caption{ (Color online) Temperature dependence of capacitance C(0)
for four different samples.} \label{Fig5}
\end{figure}
toward RT. This justifies our estimate of RT capacitances of other
four samples. However, Figure 5 reveals a weak anomaly at
temperatures close to T$_{CO}$. The model from Fig. 4 can not explain
the decrease of C(0) at T$>$T$_{CO}$. It appears that dielectric
response in PCMO is influenced by bulk properties, indeed. Let's
probe it by other methods.\vspace{0.1cm}

\emph{Measurements in magnetic field }\vspace{0.1cm}

 The influence
of magnetic field on dielectric response in PCMO is clearly one of
the most intriguing questions, having in mind colossal effects of
magnetic field on materials resistance. Figures 6a and 6b present
the influence of magnetic field on the real part of capacitance C
for T=80K and T=30K, respectively, for
single crystal sample S3.
\begin{figure}
\resizebox{0.45\textwidth}{!}{\includegraphics*{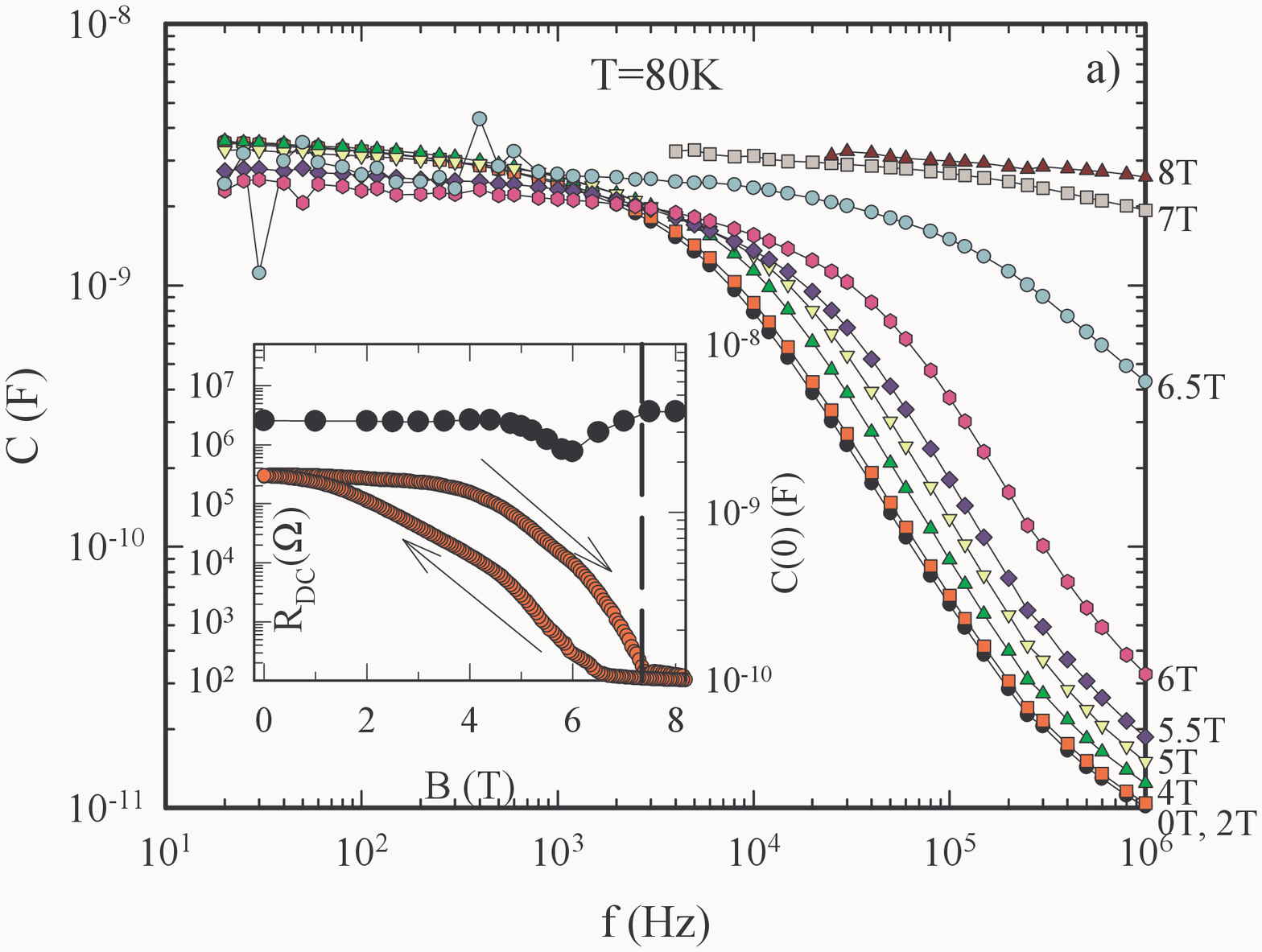}}
\resizebox{0.45\textwidth}{!}{\includegraphics*{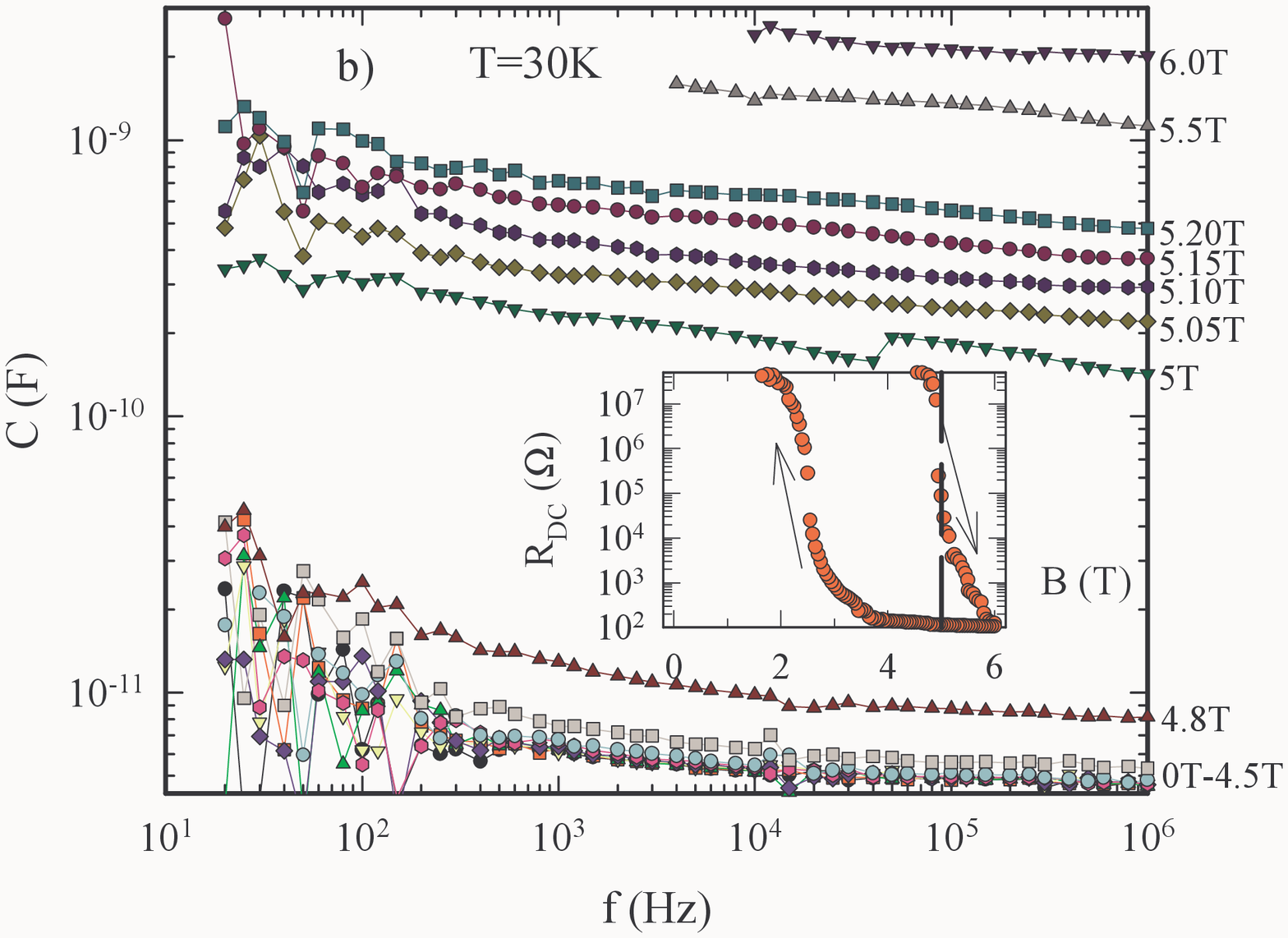}}
\caption{ (Color online) Effect of magnetic field on capacitance at
a) T=80K and b) T=30K for sample S3. Values of magnetic field are
given on the right axis. Insets show d.c. resistance as a function
of field. Broken lines indicate ferromagnetic transition B$_m$ as
deduced from up-field magnetization measurements. Solid circles in
inset of 6a denote capacitance C(0).} \label{Fig6}
\end{figure}
Insets present d.c. resistance curves in field. These insets are
excellent examples of colossal magnetoresistance effect. At field
strengths above several tesla, the resistance drops for several
orders of magnitude. This is the consequence of magnetic field
induced ferromagnetic transition \cite{Tomioka95}. Our magnetization
measurements, as well as hysteresis in resistive measurements, show
that these transitions are of first order. At T=80K, the resistance
decreases rather smoothly in magnetic field until first order
transition field of B$_{m}$=7.35T (as deduced by magnetization
measurements) that is indicated by vertical line. At B=0T resistance
is already low enough and we observe relaxation in our frequency
window. In this resistance range $\tau_0$ is given mainly by
R$_{1}$=1/G$_{1}$ and decrease of resistance enables us to follow
adjacent decrease of relaxation time $\tau_0$. At the same time,
capacitance remains roughly constant. From the data presented in
Fig. 6a we see that this case resembles very closely the ordinary,
zero-field temperature dependence (Fig. 2). Note also that the
relaxation observed in 80K case is at fields lower than B$_m$.

    Situation is different at T=30K. At 30K and zero magnetic field
dielectric relaxation falls below our frequency window (f$_0<$20Hz).
Therefore, what we see at this field is high frequency tail of our
relaxation giving high-frequency dielectric function
$\varepsilon_{HF}$$\approx$30. However, with increasing field, i.e.
decreasing resistance, one would expect to see low-frequency
relaxation reappearing as in Fig. 2a. Instead, dielectric response
increases abruptly, keeping its frequency independent (flat) shape
to the highest fields. It is difficult to explain a raise of
capacitance at constant temperature for different fields/resistances
if one would assume single element (bulk) source. But if one assumes
double capacitances like in Fig. 4, one can simulate this increase.
Modeling resistances from eq. 3 approximately yield to
C(0)$\propto$R$_2$$^2$/R$_1$$^2$. Raise of capacitance thus suggests that
contact resistance R$_2$ doesn't decrease with field as fast as bulk
resistance R$_1$. This shouldn't be unexpected if one recalls that
contact region should be the region with larger imperfections. Since
the bulk insulator-metal transition is connected with ferromagnetic
ordering, it is expectable that ferromagnetic ordering (i.e smaller
resistance) is impeded at sample boundaries, close to electrodes.
Such an interpretation is in accordance with rather high resistance
(R=R$_1$+R$_2$=100$\Omega$) above B$_m$ - this resistance is
presumably coming from contacts. In this way interplay of
resistances R$_1$=1/G$_1$ and R$_2$=1/G$_2$ give the same increase
of C(0) like in Figure 2a. The interesting phenomena is that
30K case lacks the relaxation that would be expected from gradual
decrease of R$_1$ (capacitance in Fig.6b is nearly flat in
frequency). This experimental fact is confirmed in other samples
and is always connected with ferromagnetic state at fields
above B$_m$. It seems that relatively simple equivalent diagram in Fig.4 does not represent well our system in
ferromagnetic phase. Lack of relaxation might indicate strong
correlation effects in ferromagnetic phase. Dielectric response of
systems with strong correlation is rigid, i.e. it does not relax
\cite{Jonscher}. This is exhibited by flattening/disappearance of
the loss peak and corresponding effect in its real counterpart.
Contrary to this, at T=80K we observe the relaxation since these
measurements are done in antiferromagnetic phase (B$<$B$_m$),
equally as measurements without magnetic field. Bulk and contact
resistance here depend similarly on magnetic field (see figure 1)
and C(0)$\propto$R$_2$$^2$/R$_1$$^2$ remains approximately constant. Small
twist of C(0) around B=6T can be associated exactly to the change
in resistance ratio R$_2$/R$_1$ close to ferromagnetic
transition.\vspace{0.1cm}

\emph{Measurements with d.c. bias}\vspace{0.1cm}

As already mentioned in introduction, the PCMO system is susceptible
to applied electric field, the effect that for high voltages leads
to insulator-metal transition. But even below this threshold field,
the current-voltage characteristic is nonlinear showing decrease of
resistance with increase of voltage. This nonlinearity can not be
explained solely by heating effects since at temperatures below 60K
it is observed even for heating power less than 1pW. It is therefore
interesting to see the effect of voltage on dielectric response. We
have performed dielectric measurements for a set of a.c. excitation
voltages and also those biased by d.c. voltage. They are essentially
identical so we present here just d.c. biased measurements.
\begin{figure}
\centering\includegraphics[clip,scale=0.43]{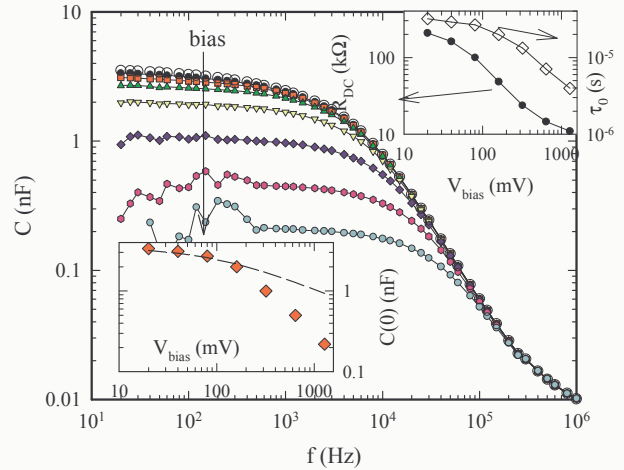}
\caption{ (Color online) Effect of d.c. bias on capacitance at T=80K
for sample S3. Arrow indicates increasing bias voltage - voltages
are as shown in the inset. Right inset: resistance R$_{DC}$
(circles, left axis) and relaxation time $\tau_0$ (diamonds, right
axis) as a function of d.c. bias voltage. Left inset: capacitance
C(0) (diamonds) vs. bias voltage. Broken line is an estimate
according to eq. 4.} \label{Fig7}
\end{figure}
Figure 7 shows such measurements for single crystal S3 at T=80K.
Capacitance is shown for a set of d.c. bias voltages (V$_{bias}$=0,
20, 40, 80, 160, 320, 640 and 1280mV). The a.c. excitation was
always kept at the lowest level of V=20mV. In the upper right inset
is the effect of bias on sample resistance (circles - left axis) and
relaxation time $\tau_0$ (diamonds - right axis). Resistance
decreases for more than one order of magnitude for this range of
voltages. Its origin is clearly not heating since resistance at this
temperature decreases even for heating power of 10nW. As in all
previous (zero-bias) cases in this study, the relaxation time
$\tau_0$ follows the resistance, i.e. decreasing resistance is
accompanied by decreasing $\tau_0$. However, we can see that this
correspondence is not perfect: $\tau_0$ decreases nine times while
resistance drops twenty-two times. If we recall our model in Fig.4,
and remember that $\tau_0$$\propto$R$_1$, this means that contact
resistance R$_2$ coming from our measured resistance R
(R=R$_1$+R$_2$) is responsible for stronger decrease of R than
expected from $\tau_0$$\propto$R$_1$. But the most important feature
here is the decrease of capacitance with bias voltage (and
decreasing resistance). This effect is opposite than observed at
temperature dependence of non-biased measurements (see for example
Fig. 3).

    It is well known that
metal-semiconductor contact usually results in Schottky barrier. A
region of semiconductor in direct contact with metal is depleted of
carriers as a consequence of different band levels in both
materials. This depletion layer thus depends on type of metal, i.e.
its energy band. The width w of depletion layer is given by

$$w=\sqrt{\frac{2\varepsilon}{qN_D}(V_b-V_a)}\eqno{(4)}$$

where V$_b$ stands for internal, junction voltage and V$_a$ for
external, applied voltage. $\varepsilon$ is dielectric constant of
semiconductor, q electron charge and N$_D$ impurity concentration.
Schottky contacts always present a capacitance C=$\varepsilon$S/w
that, as seen from eq. 4, depends as
C$\propto$w$^{-1}$$\propto$(V$_b$-V$_a$)$^{-1/2}$. In ideal Schottky
semiconductor case (one contact perfectly ohmic and another with
Schottky barrier) applied voltage V$_a$ is supposed to be in reverse
direction (negative) in order to decrease capacitance. In our case
that has two equivalent Schottky contacts, situation becomes more
complicated. But overall behavior is the same. In the left inset we
plot C(0) vs. V$_{bias}$ as deducted from Fig.7. We also plot a line
that fits to calculated data and is calculated by eq. 4 using
V$_b$=0.1V, as estimated roughly from R(T) measurement. This
estimate fits quite well to measured data for small voltages, i.e.
for voltages smaller than Schottky breakthrough voltage. The
decrease of capacitance with voltage can be therefore interpreted by
Schottky effect. And finally, current-voltage nolinearity (shown in
our case as decreasing resistance with applied voltage) is the basic
property of Schottky diodes. It thus becomes evident that dielectric
response in PCMO, as well as related nonlinear effects, has the
origin in Schottky barriers at metallic contacts. This also explains
dependence of capacitances on type of contacts (Au film or Ag
paint).\vspace{0.2cm}

    \textbf{Summary}\vspace{0.1cm}

We have performed detailed analysis of "colossal" dielectric
response in Pr$_{0.6}$Ca$_{0.4}$MnO$_3$. Dielectric relaxation and
decrease of capacitance at low temperatures are associated with the
interplay of surface and bulk capacitances and their related
resistances. Change of capacitance with magnetic field can be
equally well explained by surface (contact) capacitances. Dependence
of dielectric response on voltage (both d.c. and a.c.) can be
explained only as a consequence of Schottky layers in contact with
electrodes. This interpretation is in accordance with dependence of
capacitance on metal used as the electrode. All of the above suggest
that bulk properties of title material are not responsible for
dielectric response: dielectric constant of title material is
$\varepsilon$(0)=$\varepsilon$$_{HF}$=30, of the same order of
magnitude as in other perovskites \cite{perovskite_eps}. None of the
intriguing physical properties of PCMO (charge ordering, antiferro
and ferromagnetism, clusters) seem to influence dielectric response.
The only feature resembling to bulk property of PCMO is a weak
anomaly at temperatures close to temperature of charge ordering.
Decrease of capacitance with temperature for T$>$T$_{CO}$ can not be
explained in the frame of diagram in Fig. 4. However, it might be
evident that contact (Schottky) capacitance C$_2$ can be temperature
dependent through internal voltage V$_b$. This can lead to slight
increase of C$_2$ with lowering temperature. And at low temperatures
we enter into regime described by Fig. 4. As can be seen from
4-contacts resistance curve of sample S3 in Fig. 1, bulk resistance
of PCMO increases rapidly at T$_{CO}$. This rapidly influences the
balance of resistances in Fig. 4, resulting in decrease of overall
capacitance. Thus, T$_{CO}$ anomaly can be again interpreted as
interplay of bulk and contact capacitance.

    The temperature dependence of dielectric constant in Ref.
\cite{Mercone} generally agrees with those presented here. The most
visible difference is much stronger anomaly at T=T$_{CO}$ in Ref.
\cite{Mercone}. We interpret this by different contact material
(GaIn paint) that was used in that study. Since contacts depend on
type of metal used, it should be expected that contact capacitances
have different temperature dependence and also different breakdown
voltages. Usage of rather high voltage of V=1V in this study could
additionally influence Schottky capacitances.

    Our measurements, in accordance with some reports
\cite{Lunkenheimerfake}, suggest strongly that all reports of
apparently colossal dielectric constant should pass detailed
analysis in order to eliminate the possibility of Schottky barrier
capacitances as the origin of anomalously large dielectric constant.
As for the family of PCMO manganites, we hope that we proved such an
origin. Our finding is emphasized by apparently colossal dielectric
constant in other manganites (CuCa$_3$Mn$_2$Ti$_2$O$_{12}$) that,
contrary to PCMO, lack charge ordering or structural inhomogeneities
\cite{jaCCMTO}.\vspace{0.1cm}

    Finally, it is worth to give one more comment about the colossal effect of magnetic
field on dielectric response in title material (Fig. 6). This
"magnetocapacitive" or "magnetodielectric" effect, has recently
attracted considerable interest \cite{Kimura, Hemberger, Hur}. We
have demonstrated here that colossal magnetocapacitive effects can
also arise from purely non-intrinsic contributions.\vspace{0.2cm}

$\textbf{Acknowledgments}$\vspace{0.1cm}

We acknowledge financial support from Ministerio de
Educaci$\acute{o}$n y Ciencia (MAT2003-01880) and Comunidad de
Madrid (07N/0080/2002). We are very grateful to R. Jim$\acute{e}$nez
Riob$\acute{o}$o, for enlightning discussions.

\end{document}